\documentclass{article}
\usepackage{spconf,amsmath,graphicx,hyperref}

\usepackage{amssymb, enumitem}

\usepackage[table]{xcolor}
 
\usepackage{tabularx}
\definecolor{oursrow}{gray}{0.92}
\definecolor{best}{RGB}{200,235,200}
\newcommand{\best}[1]{\cellcolor{best}\textbf{#1}}

\let\oldbibliography\thebibliography
\renewcommand{\thebibliography}[1]{\oldbibliography{#1}\setlength{\itemsep}{0pt}}

\title{RePlay: Retrieval-Based Voice Playback for Multi-Turn spoken dialogue}
\name{Sathvik Udupa$^{1,2}$, Naveen Kumar$^{2}$, Ryan Folmsbee$^{2}$}
\address{$^{1}$BUT Speech@FIT, Czechia\\
	$^{2}$Disney Research Imagineering, Los Angeles, USA}

\begin{document}
%
\maketitle

\begin{abstract}
Many voice interaction applications require exact control over both the content and delivery of responses, typically using pre-recorded lines. Recent full-duplex models respond with low latency but cannot guarantee exact content or reproduce a specific recorded performance, while cascaded systems can be constrained to predefined responses at the cost of additional latency. We propose RePlay, a spoken dialogue system adapted from PersonaPlex that handles multi-turn conversations by retrieving and playing pre-recorded lines. Using probing, we identify the layer and frame at which the upcoming response becomes recoverable, and use this hidden state as the retrieval query. RePlay retains only the layers up to that point and replaces text and speech generation with lightweight turn-taking and retrieval heads. In simulated multi-turn interviews, RePlay reaches a median latency of 383\,ms, 3 to 7 times lower than ASR-LLM cascades of comparable dialogue quality, at the cost of lower exact-line accuracy. In a user study, participants preferred RePlay in 63\% of ratings versus 12\% for a fast cascade with a small LLM ($p=0.008$), and showed a non-significant preference (46\% vs.\ 21\%) over a slower cascade with a stronger LLM.
\end{abstract}

\vspace{-3mm}
\section{Introduction}
\label{sec:intro}
\vspace{-2mm}

Recent end-to-end spoken dialogue models process user speech continuously, respond with low latency~\cite{defossez2024moshi}, and support increasingly controllable behaviour~\cite{roy2026personaplex, tyagi2026steerduplex, chien2026moshirag}. However, their open-ended generation cannot guarantee exact content or reproduce a specific recorded performance. This limits their use where responses must follow predefined content and delivery, such as interactive characters in entertainment and games, venue guides, and customer service agents.

In such applications, the task reduces to deciding when to respond and which response to select from a known set. Cascaded systems detect the end of the user's turn \cite{udupa26_interspeech}, transcribe speech, and select a response with a traditional dialogue system or an Large Language Model (LLM), where LLM-based selection achieves higher recall at the cost of higher latency.

Full-duplex speech models are well suited for low-latency interaction because they model turn-taking internally. Text LLMs are known to represent upcoming content in their hidden states before generating it~\cite{pal2023future, dong2025emergent, pochinkov2025parascopes}. We hypothesise that full-duplex speech models similarly form a representation of their whole response at the start of the system turn, and leverage it to directly predict an embedding for retrieval from a fixed set of lines, combining the latency of end-to-end models with the control of response selection. We propose \textbf{RePlay}, which adapts PersonaPlex~[2] on this principle. Through probing, we identify the earliest frame and layer at which the upcoming response becomes recoverable, retain only the layers up to this point, and replace text and speech generation with lightweight turn-taking and retrieval heads, the latter mapping the hidden state to the embedding of the appropriate line. Turn-end detection and response selection are thus performed jointly, without Automatic Speech Recogntion (ASR) or speech synthesis.

Our contributions are:
(1) a probing analysis showing that upcoming response of a full-duplex model becomes recoverable at the start of turn;
(2) RePlay, a truncated spoken dialogue system with turn-taking and retrieval heads for low-latency response retrieval;
(3) a simulated multi-turn evaluation and a user study, showing a median latency of 383\,ms, 3 to 7 times lower than cascades of similar dialogue quality.

%


\vspace{-3mm}
\section{Related Work}
\label{sec:related}
\vspace{-2mm}
\noindent\textbf{Retrieval from speech.} Neural embedding models~\cite{reimers2019sentence} enable retrieval by semantic similarity~\cite{hambarde2023information, huang2020embedding} and extended to retrieval-augmented generation~\cite{gao2023retrieval}. Recent work bypasses ASR by embedding speech directly: SpeechRAG aligns a speech encoder to a frozen text retriever~\cite{mundnich2025speech}, and other approaches retrieve from spoken queries in dialogue~\cite{chen2025wavrag, feng2025enhancing, wang2024retrieval} or over long-form audio~\cite{someki2026planrag}. These methods retrieve knowledge to ground generation, once per query after the user has finished speaking.

\noindent\textbf{Retrieval in spoken dialogue.} Recent cascaded~\cite{udupa2026full} and end-to-end~\cite{arora2025stream, chien2026moshirag, liang2026voxmind, zhang2026duplexsla} systems hide retrieval and tool-call latency by querying while user is speaking or retrieving asynchronously during the response. However, the model must still generate a response conditioned on the retrieved content, so the retrieved information reaches the user only after further generation, and exact content and delivery cannot be guaranteed. RePlay instead retrieves the response itself at the turn boundary, so the selected line can be played immediately.

\vspace{-2mm}
\section{Proposed approach}
\label{sec:method}
\vspace{-2mm}
We first describe PersonaPlex, then use probing to identify how to adapt it for retrieval, and finally present our architecture and training.
\vspace{-2mm}
\subsection{PersonaPlex}
\label{ssec:personaplex}
PersonaPlex~\cite{roy2026personaplex} extends Moshi~\cite{defossez2024moshi} with a voice prompt for voice cloning and a text prompt for instruction following. It consists of a Mimi speech encoder, a 32-layer Transformer LLM, and a Depth Transformer that predicts the audio codebooks of the system speech, decoded to a waveform by the Mimi decoder. The LLM predicts a system text stream, fed back autoregressively, with three token types: \texttt{<WRD>} for word tokens, \texttt{<EPAD>} typically placed immediately before each word, and \texttt{<PAD>} elsewhere, including during user speech. The first \texttt{<EPAD>} after a user turn thus marks the start of the system response.

\begin{figure}
\begin{minipage}[b]{1.0\linewidth}
  \centering
  \centerline{\includegraphics[width=8cm]{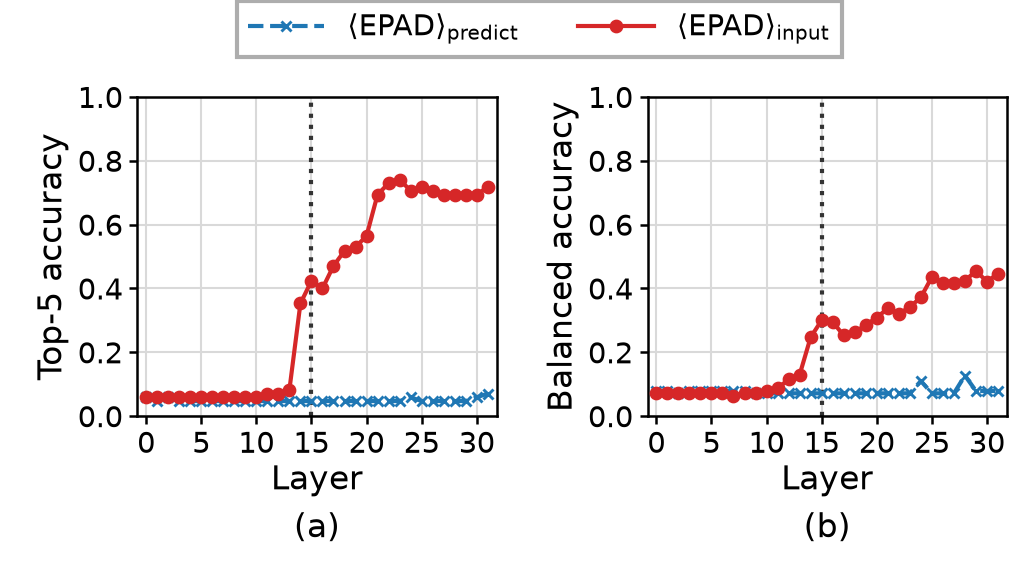}}
\end{minipage}
\vspace{-10mm}
\caption{Per-layer probing for the upcoming response: (a) QA accuracy, (b) clustering accuracy. Dotted line: layer 15.}
\vspace{-4mm}
\label{fig:probe}
\label{fig:probe}\label{fig:layers}
\end{figure}

\vspace{-2mm}
\subsection{Identifying the layer and frame for retrieval}
\label{ssec:moshi-layers}
We build on the existing capabilities of PersonaPlex, in particular its turn-taking. Since the first \texttt{<EPAD>} after a user turn marks the start of the system response, we consider two candidate frames for retrieval: the frame where this token is predicted ($\texttt{<EPAD>}_{\mathrm{predict}}$) and the next frame, where it is the input ($\texttt{<EPAD>}_{\mathrm{input}}$). The remaining question is which of the 32 layers encodes the upcoming response. Text LLMs are known to encode upcoming content in their hidden states before generating it~\cite{pal2023future, dong2025emergent, pochinkov2025parascopes}, and we test whether this holds for the spoken response. Following standard layer-wise probing~\cite{gurnee2024language}, we train a ridge regression at each layer and both frames to map the hidden state to the nomic-embed-text embedding~\cite{nussbaum2024nomic} of PersonaPlex's response, using single-turn general question answering\footnote{Our target domain is casual conversation with an interactive agent, where responses rarely contain distinctive keywords, so the representation must encode their general content rather than a specific keyword. We thus use 200/100 train/test questions with general, keyword-free answers.}. We evaluate top-5 retrieval accuracy among held-out responses and balanced accuracy over $k{=}16$ k-means clusters of response embeddings.

As Fig.~\ref{fig:layers} shows, the response is encoded only once \texttt{<EPAD>} is fed back as input. At $\texttt{<EPAD>}_{\mathrm{predict}}$, the probe stays near chance at all layers, whereas at $\texttt{<EPAD>}_{\mathrm{input}}$ it rises sharply around layers 13 to 14 and keeps improving thereafter. We therefore read the state at $\texttt{<EPAD>}_{\mathrm{input}}$ from layer 15, the earliest layer after this rise, and retain only the first 16 LLM layers, halving the LLM compute while keeping a clear response signal.

\begin{figure}[t]
\begin{minipage}[b]{1.0\linewidth}
  \centering
  \centerline{\includegraphics[width=8.5cm]{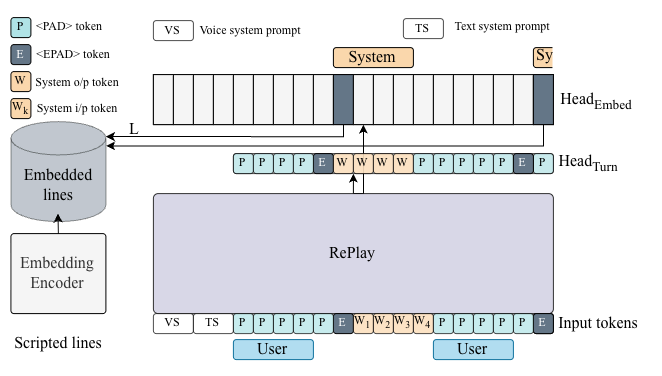}}
\end{minipage}
\vspace{-8mm}
\caption{RePlay model architecture}
\vspace{-5mm}
\label{fig:replay}
\end{figure}
\vspace{-3mm}
\subsection{RePlay model}
\label{ssec:replay}
RePlay (Fig.~\ref{fig:replay}) retains the first 16 LLM layers of PersonaPlex and discards the remaining layers along with speech generation. Instead of generating speech, it selects a scripted line and plays its pre-recorded audio. The input follows PersonaPlex, with two differences. First, \texttt{<EPAD>} is kept only before the first word of each response, so that it serves as a dedicated turn-start token. Second, played lines take the place of generated speech: their audio is fed on the system channel, and their words are inserted into the text stream at forced-aligned timings.

The text output is replaced by two linear heads on layer 15. $\mathrm{Head}_{\mathrm{Turn}}$ classifies each frame as \texttt{<PAD>}, \texttt{<EPAD>}, or \texttt{<WRD>}, where \texttt{<WRD>} groups all word tokens, and the system turn starts when \texttt{<EPAD>} is most probable. At the following frame, $\texttt{<EPAD>}_{\mathrm{input}}$, $\mathrm{Head}_{\mathrm{Embed}}$ projects the hidden state into the text embedding space and scores each line by cosine similarity to its frozen text embedding. Since no parameters are tied to specific lines, new lines can be added by embedding them, without retraining.
\vspace{-3mm}
\subsection{Retrieval training}
\label{ssec:training}
$\mathrm{Head}_{\mathrm{Embed}}$ is trained at the frame after each turn-initial \texttt{<EPAD>} to align with the frozen line embeddings~\cite{zhai2022lit}, using a one-directional SigLIP loss~\cite{zhai2023sigmoid} and a cosine alignment term:
\begin{equation}
    \mathcal{L} = -\tfrac{1}{|B|} \sum_{i \in B} \sum_{j \in \mathcal{C}(i)} \log \sigma\big(y_{ij}(\tau s_{ij} + b)\big) + \tfrac{\lambda}{|B|} \sum_{i \in B} (1 - s_{i i^{+}}),
\end{equation}
where $B$ is the set of system turns in a batch, $s_{ij}$ is the cosine similarity between projection $z_i$ and line $j$, and $i^{+}$ is the target line. $\mathcal{C}(i)$ contains $i^{+}$ and hard negatives mined by embedding similarity, excluding near-paraphrases of the target~\cite{moreira2024nv}. $y_{ij}{=}{+}1$ if $j{=}i^+$ and $-1$ otherwise, $\tau$ is a learned inverse temperature and $b$ a learned bias, and $\lambda$ weights the alignment term.

\vspace{-5mm}
\section{Experimental Setup}
\label{sec:experiments}
\vspace{-3mm}
\subsection{Datasets}
\label{ssec:data}
\noindent\textbf{Synthetic multi-character dialogue (D1).} We use an LLM ( Sonnet 5) to create 100 characters, each with a persona and about 300 deduplicated lines of at most 20 words over 30 intents (29,225 lines). This is followed by dialogue generation with free-form visitor turns, selecting character replies by line ID. We keep a quarter of visitor turns deliberately underspecified about intents, so that the model learns to rely on dialogue context and to respond with clarification lines. The 52,768 dialogues (1071\,h) are synthesised with Chatterbox TTS\footnote{\url{https://github.com/resemble-ai/chatterbox}} using cloned VCTK \cite{veaux2017vctk} and LibriSpeech \cite{panayotov2015librispeech} voices, adding long user pauses and WHAM! noise \cite{wichern2019wham}.

\begin{figure}[t]
\begin{minipage}[b]{1.0\linewidth}
  \centering
  \centerline{\includegraphics[width=6cm]{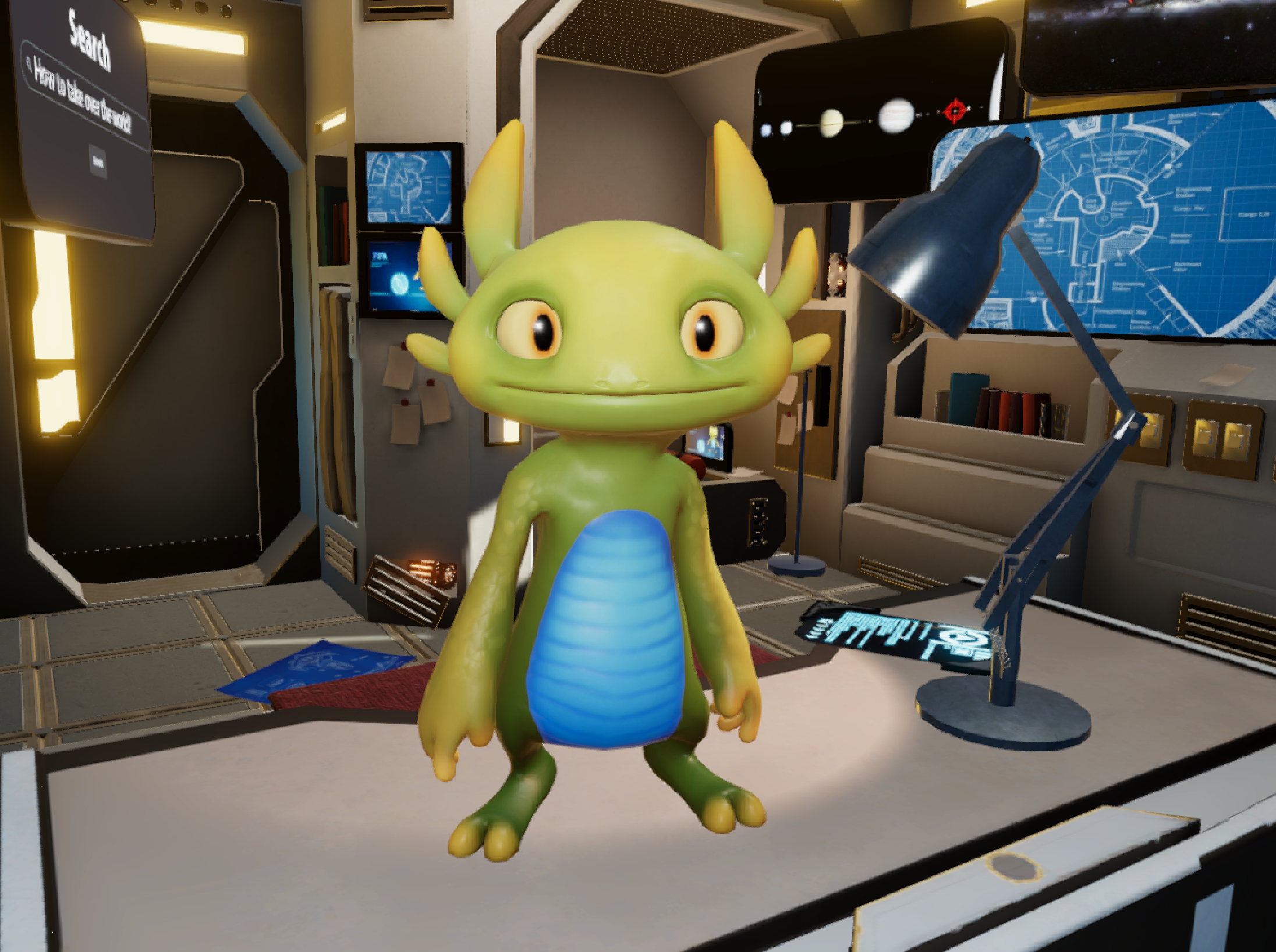}}
\end{minipage}
\vspace{-5mm}
\caption{Nobu, our agent: human-like but distinctly non-human, to keep user expectations flexible.}
\label{fig:nobu}
\vspace{-5mm}
\end{figure}

\noindent\textbf{Target interaction (D2).} As we evaluate latency, we choose a user-driven scenario: a job interview in which the user interviews Nobu \cite{paetzel2023improving}, a non-human character applying for a position (Fig \ref{fig:nobu}). We authored 300 lines for Nobu across 30 interview intents, and simulate interviews on this line set as shown for D1.

We train on D1 and then fine-tune on 2\,h of simulated D2 interviews. To test generalisation, we use three D2 fine-tuning splits: all seen (AS), where all 300 lines appear in fine-tuning; unseen lines (UL), where 50\% of the lines in each intent are held out; and unseen intents (UI), where half of the intents are held out entirely, excluding shared intents such as greeting and farewell. Although both datasets are synthetic, RePlay inherits turn-taking from PersonaPlex, whose released checkpoint is trained on real conversations alongside synthetic ones\footnote{We also tried mixing real dyadic corpora into training to improve generalisation and expose the model to real speech, but observed no benefit.}, and we test it with real users in Sec.~\ref{ssec:userstudy}.
\vspace{-3mm}
\subsection{Training details}
\vspace{-2mm}
Both heads are trained jointly with teacher forcing. As our training data is synthetic, we preserve PersonaPlex's robustness to real speech by applying LoRA (rank 256) to layers 0--9 and fully fine-tuning only layers 10--15. $\mathrm{Head}_{\mathrm{Turn}}$ uses cross-entropy with \texttt{<EPAD>} weighted by 9, and $\tau$ and $b$ are initialised at 10 and $-10$. Lines are embedded with pretrained nomic-embed-text~\cite{nussbaum2024nomic}. $\mathcal{C}(i)$ contains all D1 lines and 600 negatives shared across the batch, half from the nearest neighbours of the batch targets and half uniform, excluding lines with cosine similarity above 0.9 to the target. Training uses sequences of up to 90\,s on 8 A100 (40\,GB) GPUs.


\vspace{-4mm}
\subsection{Baselines}
\vspace{-2mm}
\label{ssec:baselines}
No existing end-to-end system, open-source or commercial, supports playing pre-recorded lines. Thus, we compare against cascaded systems combining streaming ASR and an LLM (Table~\ref{tab:baselines}). All baselines select a line at the detected end of the user turn and play its pre-recorded audio immediately. Cascade A uses a commercial LLM, while Cascades B and C use local LLMs served by vLLM~\cite{kwon2023efficient}, with partial ASR transcripts prefilling the LLM cache. Cascades A and B are prompted with Nobu's persona and all 300 lines, whereas Cascade C trades accuracy for latency, prompting a small LLM with a short persona and Nobu's topics and mapping its open-ended reply to the nearest line by embedding similarity. As in RePlay, lines are embedded without metadata.

\begin{table}[t]
    \centering
    \footnotesize
    \setlength{\tabcolsep}{3pt}
    \begin{tabularx}{\columnwidth}{|c|l|l|X|}
    \hline
    \textbf{\#} & \textbf{ASR} & \textbf{Turn end} & \textbf{Response selection} \\
    \hline
    A & Azure & ASR final & GPT-4.1 picks from the full line set \\
    \hline
    B & Kyutai~\cite{zeghidour2025streaming} & VAD $\geq 0.6$ & Gemma-4-E4B picks from the full line set \\
    \hline
    C & Kyutai~\cite{zeghidour2025streaming} & VAD $\geq 0.6$ & Gemma-3-1B free reply, mapped to the nearest line (nomic-embed~\cite{nussbaum2024nomic}) \\
    \hline
    \end{tabularx}
    \vspace{-3mm}
    \caption{Baseline systems.}
    \label{tab:baselines}
    \vspace{-5mm}
\end{table}

\begin{table*}[t]
\centering
\small
\caption{Online evaluation with a simulated interviewer (100 dialogues per system). Metrics as in Sec.~\ref{ssec:evaluation}; Q: judge rating (1--5) for engagement and responsiveness; $\pm$: std.\ over dialogues. Latency excludes cutoff turns. Best per column highlighted.}
\label{tab:online}
\begin{tabular}{l|ccc|cc|ccccc}
\hline
 & \multicolumn{3}{c|}{Retrieval} & \multicolumn{2}{c|}{Dialogue quality} & \multicolumn{5}{c}{Turn-taking} \\
 & & & & & & & \multicolumn{4}{c}{Latency (ms)} \\
System & R@1\,$\uparrow$ & R@3\,$\uparrow$ & R@10\,$\uparrow$ & Q\,$\uparrow$ & IntentAgr\,$\uparrow$ & CR (\%)\,$\downarrow$ & p10\,$\downarrow$ & p50\,$\downarrow$ & p90\,$\downarrow$ & mean\,$\downarrow$ \\\hline
Cascade A & \best{0.52} & \best{0.62} & \best{0.71} & \best{3.90}$\pm$0.33 & \best{0.85}$\pm$0.16 & \best{0.2} & 2303 & 2612 & 3012 & 2652 \\
Cascade B & 0.36 & 0.55 & 0.62 & 3.77$\pm$0.45 & 0.77$\pm$0.19 & 8.7 & 780 & 1226 & 1414 & 1145 \\
Cascade C & 0.17 & 0.27 & 0.34 & 2.99$\pm$0.63 & 0.35$\pm$0.19 & 17.0 & \best{296} & 591 & 784 & 569 \\
\rowcolor{oursrow}
RePlay (ours) & 0.35 & 0.50 & 0.60 & 3.76$\pm$0.43 & 0.83$\pm$0.16 & 3.6 & 333 & \best{383} & \best{439} & \best{387} \\
\hline

\end{tabular}
\vspace{-5mm}
\end{table*}

\vspace{-3mm}
\subsection{Evaluation}
\label{ssec:evaluation}
\vspace{-2mm}

We simulate the interviewer with an LLM (GPT-4o-mini), prompted with a persona, an interview goal, and an ordered list of high-level topics, but without access to the system's intents or lines, so it cannot steer towards particular responses. User turns are synthesised with Breeze-TTS-2\footnote{\url{https://huggingface.co/BreezeBlue/Breeze-TTS-2}} using a per-dialogue cloned voice, trimmed with forced alignment~\cite{mcauliffe2017montreal}, and inserted 0.4\,s after the system finishes speaking. We run 100 scripted scenarios of up to 90\,s each. In all evaluations, each system acts as Nobu, selecting from his 300 lines (D2) embedded with nomic-embed-text~\cite{nussbaum2024nomic}. 

Since the simulated dialogues have no reference lines, an LLM judge (GPT-4o-mini), given the full dialogue and all lines grouped by intent, selects the intent each system turn should answer and ranks its three best lines. We report intent agreement (IntentAgr), the fraction of played lines within the judge's intent, and recall at $k$ (R@$k$), where the reference set is the judge's top line ($k{=}1$), its three ranked lines ($k{=}3$), or these extended with the seven lines most similar to them under nomic-embed-text\cite{nussbaum2024nomic} ($k{=}10$). The judge also rates each dialogue from 1 to 5 for engagement and responsiveness.

For turn-taking, we report the 10th, 50th, and 90th percentiles and mean latency from the end of user speech to the system reply. Replies issued before the user finishes are reported separately as the cutoff rate (CR, \% of turns).
\vspace{-4mm}
\subsection{User study}
\label{ssec:userstudy}
\vspace{-3mm}
In user study, $N{=}14$ participants each interviewed the alien agent as RePlay and one baseline, in counterbalanced order\footnote{Cascade A was excluded due to its substantially higher latency.}. They received only a brief scenario description, with no intents or example questions, so their questions were open-ended within the domain. Sessions lasted 78--253\,s (median 149\,s), followed by a survey on quality and responsiveness.

\vspace{-2mm}

\begin{table}[b]
\centering
\footnotesize
\setlength{\tabcolsep}{3.5pt}
\vspace{-8mm}
\caption{Ablation on training data. D1: synthetic pretraining. D2: 2\,h fine-tuning on Nobu with all lines seen (AS), held-out lines (UL), or held-out intents (UI). Metrics as in Table~\ref{tab:online}.}
\label{tab:ablation}
\begin{tabular}{cc|ccc|cc}
\hline
D1 & D2 & R@1 & R@3 & R@10 & Q & IntentAgr \\
\hline
\multicolumn{7}{l}{\textit{Training stages}} \\
\checkmark & -- & 0.18 & 0.28 & 0.37 & 3.16$\pm$0.53 & 0.45$\pm$0.18 \\
-- & AS & 0.23 & 0.43 & 0.54 & 3.72$\pm$0.45 & 0.77$\pm$0.17 \\
\checkmark & AS & \textbf{0.35} & \textbf{0.50} & \textbf{0.60} & \textbf{3.76}$\pm$0.43 & \textbf{0.83}$\pm$0.16 \\
\hline
\multicolumn{7}{l}{\textit{Generalisation (with D1)}} \\
\checkmark & UL & 0.28 & 0.46 & 0.57 & 3.74$\pm$0.44 & 0.76$\pm$0.17 \\
\checkmark & UI & 0.21 & 0.31 & 0.35 & 3.47$\pm$0.52 & 0.65$\pm$0.18 \\
\hline
\end{tabular}
\end{table}
\vspace{-3mm}
\section{Results}
\label{sec:results}
\vspace{-2mm}

\noindent\textbf{Simulated multi-turn evaluation.} Table~\ref{tab:online} shows the trade-off between response quality and latency. Cascade A, with a large LLM selecting from the full line set, gives the best response selection and dialogue quality, but at a median latency of 2.6\,s. RePlay reduces this to 383\,ms, 6.8$\times$ and 3.2$\times$ faster than Cascades A and B, with similar dialogue quality (Q of 3.76 vs.\ 3.90 and 3.77) and intent agreement close to Cascade A (0.83 vs.\ 0.85). Its latency is also consistent, with only 106\,ms between the 10th and 90th percentiles, as retrieval happens at turn start rather than after ASR and LLM decoding. Cascade C approaches this latency with a small LLM, but at much lower intent agreement (0.35) and Q (2.99). RePlay also rarely interrupts the user (CR of 3.6\%), as its turn-taking head does not trigger on short pauses, whereas the VAD of Cascades B and C triggers early in 8.7\% and 17.0\% of turns.\footnote{The VAD triggers on many mid-turn pauses in both cascades. Cascade C's reply is often ready before the user resumes speaking and is played, while Cascade B's slower reply is dropped once the user continues.} The remaining gap to Cascade A lies mainly in exact line choice (R@1 of 0.35 vs.\ 0.52), suggesting that RePlay usually finds an appropriate response even when it differs from the judge's preferred line.

\noindent\textbf{Ablation.} Table~\ref{tab:ablation} shows that both training stages contribute. Without seeing Nobu D2 data, the D1-only model selects a line from the correct intent in 45\% of turns, so retrieval transfers to an unseen character to some extent due to 100-character training. Fine-tuning on D2 alone is considerably stronger, and combining both stages is best on every metric, raising R@1 from 0.23 to 0.35. For unseen lines within seen intents (UL), performance stays close to the all-seen model (Q of 3.74 vs.\ 3.76), suggesting that new lines can be added to an existing intent by embedding their text. Unseen intents (UI) are harder, with intent agreement falling to 0.65 and R@10 to 0.35. The UI model still outperforms the D1-only model in intent agreement and Q, but its gap to the all-seen (AS) model shows that new intents benefit from their own training data.

\noindent\textbf{User study.} Participants rated which conversation was better on wait time, pace, back-and-forth, and responsiveness. Against Cascade C ($n{=}8$), RePlay was preferred in 63\% of answers vs.\ 12\%, and all participants favoured it overall (Wilcoxon signed-rank, $p{=}0.008$). Against Cascade B ($n{=}6$), the preference was weaker and not significant (46\% vs.\ 21\%, $p{=}0.50$): participants favoured RePlay's shorter waits and pace, but found the baseline's replies slightly more appropriate. Presentation order had no significant effect.

\vspace{-5mm}

\section{Conclusion}
\vspace{-3mm}
\label{sec:conclusion}

We presented RePlay, a retrieval-based spoken dialogue system that selects pre-recorded lines from the hidden state of a full-duplex speech model. Probing PersonaPlex showed that the upcoming response becomes recoverable once the turn-start token is fed back, allowing us to keep only its first 16 layers and replace text and speech generation with lightweight turn-taking and retrieval heads. In simulated interviews, RePlay reached a median latency of 383\,ms, 3 to 7 times faster than cascades of comparable dialogue quality, and was preferred over a fast cascade in a user study. It generalises to new lines within seen intents without retraining, while unseen intents benefit from in-domain data and exact line selection trails a large LLM. Our evaluation relies on synthetic training data, an LLM judge, and a small user study; future work will explore retrieval with finer-grained response metadata.

\noindent\textbf{Generative AI disclosure.} Claude Opus 5 was used for light editing and grammatical improvements of the manuscript, and, via Claude Code, as a coding assistant. All AI-assisted text and code were reviewed and verified by the authors.
\\
\noindent\textbf{Compliance with ethical standards.} User-study participants gave informed consent; no personally identifiable information was collected, and it was anonymised.

\label{sec:refs}

\bibliographystyle{IEEEbib}
\bibliography{strings,refs}

\end{document}